\documentclass[twocolumn,english,twocolumn,english,reprint,aps,prl]{revtex4-2}
\usepackage[T2A,T1]{fontenc}
\usepackage[latin9]{inputenc}
\usepackage{float}
\usepackage{amsmath}
\usepackage{amssymb}
\usepackage{graphicx}

\makeatletter

\DeclareRobustCommand{\cyrtext}{%
  \fontencoding{T2A}\selectfont\def\encodingdefault{T2A}}
\DeclareRobustCommand{\textcyr}[1]{\leavevmode{\cyrtext #1}}

\usepackage[colorlinks,citecolor=red,urlcolor=blue,bookmarks=false,hypertexnames=true]{hyperref}
\usepackage{mathtools, nccmath}
\usepackage{babel}

\makeatother

\usepackage{babel}
\begin{document}
\title{Integrable-to-Thermalizing Crossover in Non-Equilibrium Superconductors}
\author{Andrey Grankin, Victor Galitski}
\affiliation{Joint Quantum Institute, University of Maryland, College Park, MD
20742, USA}
\affiliation{Department of Physics, University of Maryland, College Park, MD 20742,
USA}
\begin{abstract}
Motivated by the experiment by M. Budden {\em et al.} [Nature Physics {\bf 17}, 611 (2021)], who observed signatures of long-lived photo-induced superconductivity, we develop an accurate analytical/computational approach to non-equilibrium superconductivity following a quench. We consider the BCS-Holstein model, which includes both integrable local electron-electron interactions and integrability-breaking electron-phonon coupling. We develop Keldysh-Eliashberg theory on the Kadanoff-Baym contour, which enables us to describe non-equilibrium dynamics of the superconductor. We consider a quench in interactions, which results in a dynamic transition from the initial superconducting state to a normal thermal state in the end of the evolution. It is shown that the dynamics contain two stages: The early-time integrable behavior, involving coherent oscillations of the superconducting order parameter, crosses over to the late-time ergodic dynamics exhibiting a thermal decay into an equilibrium state. In the former regime, our computational approach  both reproduces exact analytical results on the integrable dynamics of the order parameter (obtained for the zero-temperature initial state) and generalizes those to the case of an initial thermal state. The method also succeeds for the first time in describing both integrable-to-thermalizing crossover and the late-time thermal decay, which is shown to be consistent with the  time-dependent Ginzburg-Landau theory (with the exponential decay time dependent on the density of quasiparticle excitations). We observe the electron distribution function approaching the Fermi-Dirac thermal distribution at final stages. The details of two-time non-equilibrium dynamics depend on the density of quasiparticles in the initial state and the integrability-breaking parameters, which under certain conditions may result in a long-lived transient superconductivity consistent with experiment. 

\end{abstract}
\maketitle

\paragraph{Introduction}

A detailed understanding of thermalization in electronic systems represents an open, long-standing problem in condensed matter physics.
Thermalization mechanisms are not only of fundamental interest \citep{RDO08,S94, DKC21} but also have practical implications in the design and operation of
quantum devices~\citep{LGL05,AKM04}. A paradigmatic scenario for studying relaxation is represented by quench dynamics, where the system is initially prepared in an equilibrium
state and then one of the parameters is rapidly changed, resulting
in non-adiabatic dynamics and ultimately an approach to a new equilibrium or steady state \citep{M18, KSM15}. In this context integrable
systems represent a special class of models that exhibit a rich variety of
non-equilibrium behaviors~\citep{CDY16,BPC18}. However, integrability is fragile and is broken in real experiment resulting in ergodic dynamics. Understanding integrability breaking is another open problem \citep{SSY19,BEG15} of broad interest. 

Recent experiments in optical manipulation of superconductivity \citep{C18,MCN16}
demonstrated signatures of superconducting correlations at temperatures
substantially above the transition temperature. While the observed
effect is believed to be associated with the phonon-induced pairing
mechanisms \citep{BGB21,RYB23,CEK23, BKM17,MTE17,SKG16,GHG21}, the superconducting correlations
were observed to persist for much longer than the lifetime of phonon
excitations. This long-lived transient superconductivity is currently not understood. Our study is motivated by this experiment, but has broader impact to non-equilibrium superconductivity in both solid-state systems and cold atoms.

Previous theoretical studies of quenched superconductivity included analytical treatment of integrable models  \citep{YDG15,YD06}, which demonstrated coherent bosonic dynamics of the order parameter that was shown to  either approach a finite or zero value depending on the quench parameters. However, these behaviors did not include any pair-breaking mechanisms and consequently did not explore thermalization. The integrable methods are not generalizable to explore ergodic dynamics and other methods have to be employed. 

In this letter, we develop a generic theory of non-equilibrium superconductivity following a quench in interactions that captures both the integrable (coherent bosonic) dynamics and ergodic  (fermionic or pair-breaking) behavior including the ultimate thermalization into the canonical Fermi-Dirac state. We first explore the integrable dynamics, benchmark our numerical technique against known analytical solutions, generalize those to finite-temperature initial states and  demonstrate the persistence of  steady state superconductivity (which is present even if the initial temperature exceeds the nominal critical transition temperature after the quench). We then include the pair-breaking due to phonon scattering, which is shown to lead to thermalization followed by a Ginzburg-Landau-type relaxation of the superconducting order
parameter towards its equilibrium value. Specifically, we demonstrate
that thermalization  is accompanied by the decay of uncondensed
Cooper pairs into thermal quasiparticles. Our method is non-perturbative in both the quench speed and  electron-phonon coupling strength, which renders the conventional kinetic-equation approaches \citep{CS77} inapplicable. Moreover,
we demonstrate that the proper quantum kinetic equation \citep{K23} must
include a matrix-valued distribution function, which treats the thermal
quasiparticles and uncondensed Cooper pairs on an equal footing.

We consider the BCS-Holstein model, which contains two electron-electron interaction terms: the instantaneous contact BCS pairing term
parameterized by $\lambda$, which may be associated with  a  high-energy phonon band and a low-energy phonon band, which induces pair-breaking and inelastic scattering. At time $t=0$, the former interaction constant is quenched to lower values leading to a non-equilibrium state that undergoes a relaxational dynamics due  to the phonon bath. The Hamiltonian of the system reads $H\left(t\right)=H_{0}\left(t\right)+H_{\text{el}-\text{ph}}$:

\begin{align}
H_{0}=\sum_{k,\sigma}\xi_{k}\psi_{{\bf k},\sigma}^{\dagger}\psi_{{\bf k},\sigma}-\lambda\left(t\right)\nu_{0}^{-1}\sum_{{\bf q}}\varrho_{{\bf q}}\varrho_{-{\bf q}},\label{eq:Hinst}\\
H_{\text{el-ph}}=\sum_{{\bf q}}g_{{\bf q}}\phi_{{\bf q}}\varrho_{-{\bf q}}+\sum_{{\bf q}}\omega_{{\bf q}}a_{{\bf q}}^{\dagger}a_{{\bf q}},
\end{align}
  where $\varrho_{{\bf q}}=V^{-1/2}\sum_{{\bf k},\sigma}\psi_{{\bf k}+{\bf q},\sigma}^{\dagger}\psi_{{\bf k},\sigma}$,
$\phi_{{\bf q}}=a_{{\bf q}}+a_{-{\bf q}}^{\dagger}$, $\psi_{{\bf k},\sigma}$($\psi_{{\bf k},\sigma}^{\dagger}$)
and $a_{{\bf q}}(a_{{\bf q}}^{\dagger})$ respectively denote the
electron  and phonon annihilation (creation) operators with $\xi_{k}$ and $\omega_{q}$
being their dispersion relations. $g$ is the electron-phonon coupling strength
and $V$ is the volume of the system, $\nu_{0}$ is the electron density
of states at the Fermi level, $\lambda$ is the strength of short-range
contact interaction. 

\begin{figure}
\begin{centering}

\includegraphics[scale=0.19]{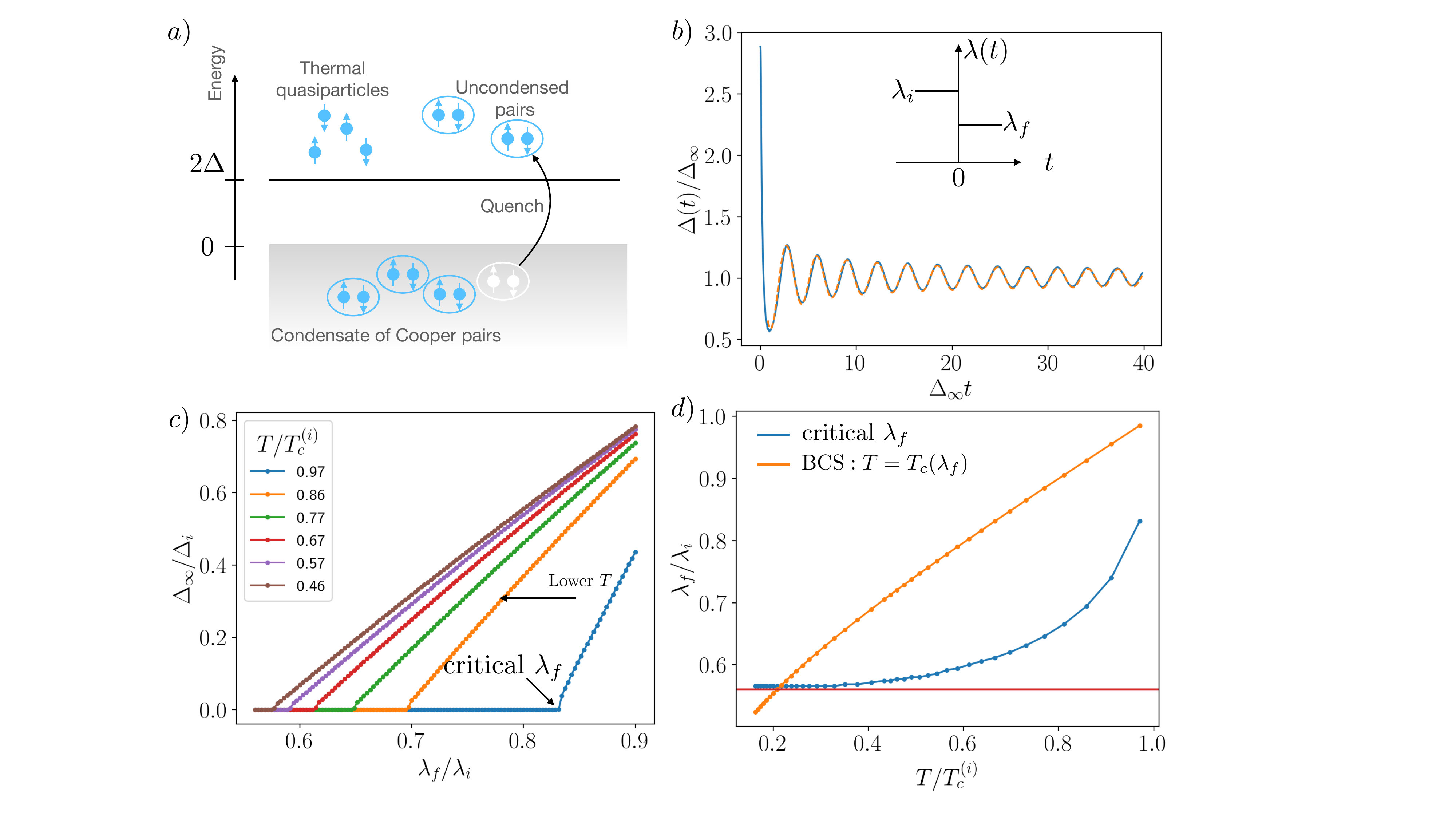}
\par\end{centering}
\caption{Order parameter dynamics after the interaction quench in a superconductor.
a) Schematic representation of the physical processes inside the quenched
superconductor at finite temperature: rapid change in the interaction
strength generates a gas of uncondensed Cooper pairs which then undergoes
an oscillatory-like dynamics. b) Dynamics of the order parameter $\Delta\left(t\right)$
for quench parameters $\lambda_{i}=0.5,$ $\lambda_{f}=0.3$ and the
initial temperature $T^{(i)}=T_{c}^{(i)}/2$. The numerical result
is shown in blue and the fit with analytical approximation (see text) is shown in dashed orange. c) Large-time behavior of the order
parameter as function of the quench parameters starting at different
initial temperatures. The critical value of the final interaction
strength is identified by the $\Delta_{\infty}\rightarrow0$. d) Critical
value of the final pairing strength as function of the initial temperature
of the superconductor. We note that the asymptotic value of $\lambda_{f}/\lambda_{i}$
at low temperatures is in a good agreement with \citep{YD06} (the
corresponding limit is shown as a red horizontal line). The orange curve represents BCS equilibrium expression of the coupling strength at which the temperature $T$ is critical. }
\label{Fig1}
\end{figure}

\paragraph{Turning off phonons --} We start our analysis assuming no electron-phonon coupling, i.e. $g_{{\bf q}}=0$. We consider
the quench of the coupling constant $\lambda=\lambda\left(t\right)=\lambda_{i}\theta\left(-t\right)+\lambda_{f}\theta\left(t\right)$,
where $\theta\left(t\right)$ is the Heaviside step-function, and
assume the system is initially prepared in a superconducting state (either thermal or ground state associated with $\lambda_i$). We use the mean-field approximation and introduce Nambu spinors $\Psi_{{\bf k}}=(\hat{\psi}_{{\bf k},\downarrow},\hat{\psi}_{-{\bf k},\uparrow}^{\dagger})^{T}$.
In this case, Eq.~(\ref{eq:Hinst}) reduces to the
conventional BdG Hamiltonian
$$
H_{\text{BdG}} =\sum_{{\bf k}}\Psi_{{\bf k}}^{\dagger}\left\{ \xi_{k}\tau_{3}+\Delta\left(t\right)\tau_{1}\right\} \Psi_{{\bf k}}=\sum_{{\bf k}}\vec{h}_{k}\vec{\sigma}^{(k)},
$$
where $\sigma_{i}^{\left(k\right)}\equiv\sum_{{\bf k}}\Psi_{{\bf k}}^{\dagger}\tau_{i}\Psi_{{\bf k}}$
and $\vec{h}_{k}\left(t\right)=\left\{ \Delta\left(t\right),0,\xi_{k}\right\} $.
The gap $\Delta$ obeys the self-consistency equation
$\Delta\left(t\right)=-\frac{1}{2}\lambda\left(t\right)\nu_{0}^{-1}\sum_{k}\langle\sigma_{1}^{(k)}(t)\rangle$.
Note that while the Anderson pseudospin operators $\sigma$ obey the canonical commutation
relations $\left[\sigma_{i},\sigma_{j}\right]=2i\epsilon_{i,j,l}\sigma_{l}$,
the Hilbert space is different from that of a spin operator at finite
temperature, $T$, due to the presence of fermions. The relevant Hesienberg equations of motion are $\frac{d}{dt}\langle\vec{\sigma}^{(k)}(t)\rangle=-2\vec{h}_{k}\times\langle\vec{\sigma}^{(k)}(t)\rangle$,
where the average is taken with respect to the initial thermal state with
$\beta=1/T$. The initial conditions for the $\sigma$ operators
which can be  found from $H_{\text{BCS}}^{\text{mf}}$:
$\langle\sigma_{3}^{(k)}(0^{-})\rangle=-\xi_{k}\tanh[\beta\Omega_{k}/2]/(2\Omega_{k})$
and $\langle\sigma_{1}^{(k)}(0^{-})\rangle=-\Delta_{i}\tanh[\beta\Omega_{k}/2]/(2\Omega_{k})$,
$\Omega_{k}=\sqrt{\Delta_{i}^{2}+\xi_{k}^{2}}$ and $\Delta_{i}$
is the thermal value of the gap corresponding to $\lambda_{i}$. We
emphasize that the time-evolution associated with $H_{\text{BdG}}^{\text{mf}}$ is integrable and does not break the existing Cooper pairs (see Fig.~(\ref{Fig1})~(\textcyr{\cyra})).

As was shown in \citep{YD06}, for a quench from the ground state ($\beta=\infty$)
the gap at large times $\Delta_{\infty}\equiv\Delta\left(t\rightarrow\infty\right)$
can either approach a constant value or decay to zero depending on the integrals of motion and the values of $\lambda_{i}$ and $\lambda_{f}$. Here we extend the analysis
to finite temperatures by iteratively solving the set of differential equations with the thermal initial conditions. To this
end, we assume a constant density of single-particle states within $\left[-E_{0},E_{0}\right]$, where $E_{0}$ is  an energy cut-off. The results of numerical simulation are shown
in Fig.~\ref{Fig1}~(b). We  fit the numerical curve at large
times with the analytical solution \citep{YDG15,YD06} $\Delta\left(t\right)\approx\Delta_{\infty}\left[1+a\cos\left(2\Delta_{\infty}t+\pi/4\right)/\sqrt{\Delta_{\infty}t}\right]$
describing the amplitude oscillation with the two fitting parameters
$a$ and $\Delta_{\infty}$. The fit at large times
is displayed in Fig.~\ref{Fig1}~(b) and shows an excellent
agreement between the two methods. This  procedure also allows us to extract the large-time
integrable behavior (e.g.,  the final value of the order parameter) from numerical data for a wide range of quench parameters.
In the phonon-less regime, we generally reproduce dynamics of the order parameter  consistent with the analytical studies of the integrable Richardson model~\citep{YD06,YDG15}. 

Thermal fluctuations increase the critical value of $\lambda_{f}$
for which the terminal value of the order parameter $\Delta_{\infty}$ is finite, as shown in Fig.~\ref{Fig1}~(c). Remarkably, we find that the non-equilibrium $\Delta_{\infty}$ can
remain finite even if the equilibrium value of the order parameter corresponding to $\lambda_{f}$ is zero as shown in Fig.~\ref{Fig1}~(d). The effect is found to be most prominent if the initial temperature sufficiently close to the BCS superconducting transition temperature $T_{c}^{(i)}\equiv T_c(\lambda_i)$ before the
quench. At very low temperatures, the critical value of $\lambda_{f}$
saturates to $\lambda_{f}/\lambda_{i}=1/\left(1+\pi\lambda_{i}/2\right)$ in accordance with Ref.~\citep{YD06}. 

\paragraph{Phonon-induced relaxation ---}
We now discuss the phonon-induced relaxation of the order parameter
\begin{equation}
\left(i\partial_{t}-\hat{h}_{k}\left(t\right)-\hat{\Sigma}_{k}\star\right)\hat{G}_{k}\left(t,t'\right)=\delta\left(t,t'\right)\hat{\mathbb{I}},\label{eq:G}
\end{equation}
\useshortskip where $\hat{G}_{k}$ is the contour-ordered Green's
function defined as $\hat{G}_{k}\left(t,t'\right)=-i\left\langle T_{{\cal C}}\Psi_{k}\left(t\right)\otimes\Psi_{k}^{\dagger}\left(t'\right)\right\rangle $,
$\hat{h}_{k}\left(t\right)=\xi_{k}\hat{\tau}_{3}+\Delta\left(t\right)\hat{\tau}_{1}$
denotes the BdG Hamiltonian $\delta$ denotes the contour Dirac delta
function, and $\star$ denotes the matrix multiplication in temporal
and Nambu indices. $\hat{\Sigma}$ is the contour self-energy determined
via Dyson's equation: \useshortskip
\begin{equation}
\hat{\Sigma}_{{\bf k}}\left(t,t'\right)=-i\frac{1}{V}\sum_{k}g^{2}D_{{\bf k-k'}}\left(t,t'\right)\hat{\tau}_{3}\hat{G}_{{\bf k'}}\left(t,t'\right)\hat{\tau}_{3},\label{eq:sigma}
\end{equation}
\useshortskip where the contour-ordered phonon propagator is defined
as $D_{{\bf q}}\left(t,t'\right)=-i\left\langle T_{{\cal C}}\phi_{{\bf q}}\left(t\right)\phi_{-{\bf q}}\left(t'\right)\right\rangle $.
In the following we ignore the back-action effects on the phonon propagator and assume it is in equilibrium state at the initial temperature. We also ignore Coulomb repulsion \citep{GG23}  assuming that its effects are included in the effective BCS pairing strength. We then proceed with the standard quasiclassical approximation and average the propagator over the Fermi surface $g_{{\bf q}}^{2}D_{{\bf q}}\left(t,t'\right)\rightarrow\left\langle g_{{\bf q}}^{2}D_{{\bf q}}\left(t,t'\right)\right\rangle {}_{\text{FS}}$.
Under these assumptions, the effective phonon propagator is fully determined
by its spectral density (Eliashberg function) which we denote as $\alpha^{2}F\left(\omega\right)\equiv\pi^{-1}2\nu_{0}\text{Im}\left\langle g_{{\bf q}}^{2}D_{{\bf q}}^{R}\left(\omega\right)\right\rangle {}_{\text{FS}}$,
where the $D_{{\bf q}}^{R}\left(\omega\right)$ is the retarded
part phonon propagator. The effective electron-phonon
pairing strength and the mean frequency are conventionally defined
as $\lambda_{\text{el-ph}}=\int\frac{d\omega}{\omega}\alpha^{2}F\left(\omega\right)$,
$\bar{\omega}=\lambda_{\text{el-ph}}^{-1}\int d\omega\alpha^{2}F\left(\omega\right)$.
The time-dependent order parameter, $\Delta\left(t\right)$, satisfies the self-consistency equation by replacing the phonon propagator in
Eq.~(\ref{eq:sigma}) with $\lambda\nu_{0}^{-1}\delta\left(t,t'\right)$.
For concreteness, we follow \citep{KCL76} and assume the following model of acoustic phonons  $\alpha^{2}F\left(\omega\right)=2\lambda_{\text{el-ph}}\left(\omega/\omega_{D}\right)^{2}$$\theta\left(\omega_{D}-\omega\right)$,
where $\omega_{D}$ is  the Debye frequency cut-off, which yields
$\bar{\omega}=2\omega_{D}/3$. In the following, we assume $\bar{\omega}/T\ll1$.
In this case, the phonons do not dominate the electron
pairing and mostly serve as a relaxation mechanism. The set of equations
Eqs.~(\ref{eq:G}, \ref{eq:sigma}) can be solved numerically by
discretizing both the time variable on the Kadanoff-Baym contour  and the energy-momentum variables. In the current work, we use adapted routines from the open-source software available at \citep{SGM20} and develop our own specialized non-equilibrium superconductivity (non-equilibrium superconductivity; NESc) code, which we open source as well. 

\begin{figure}
\begin{centering}
\includegraphics[scale=0.22]{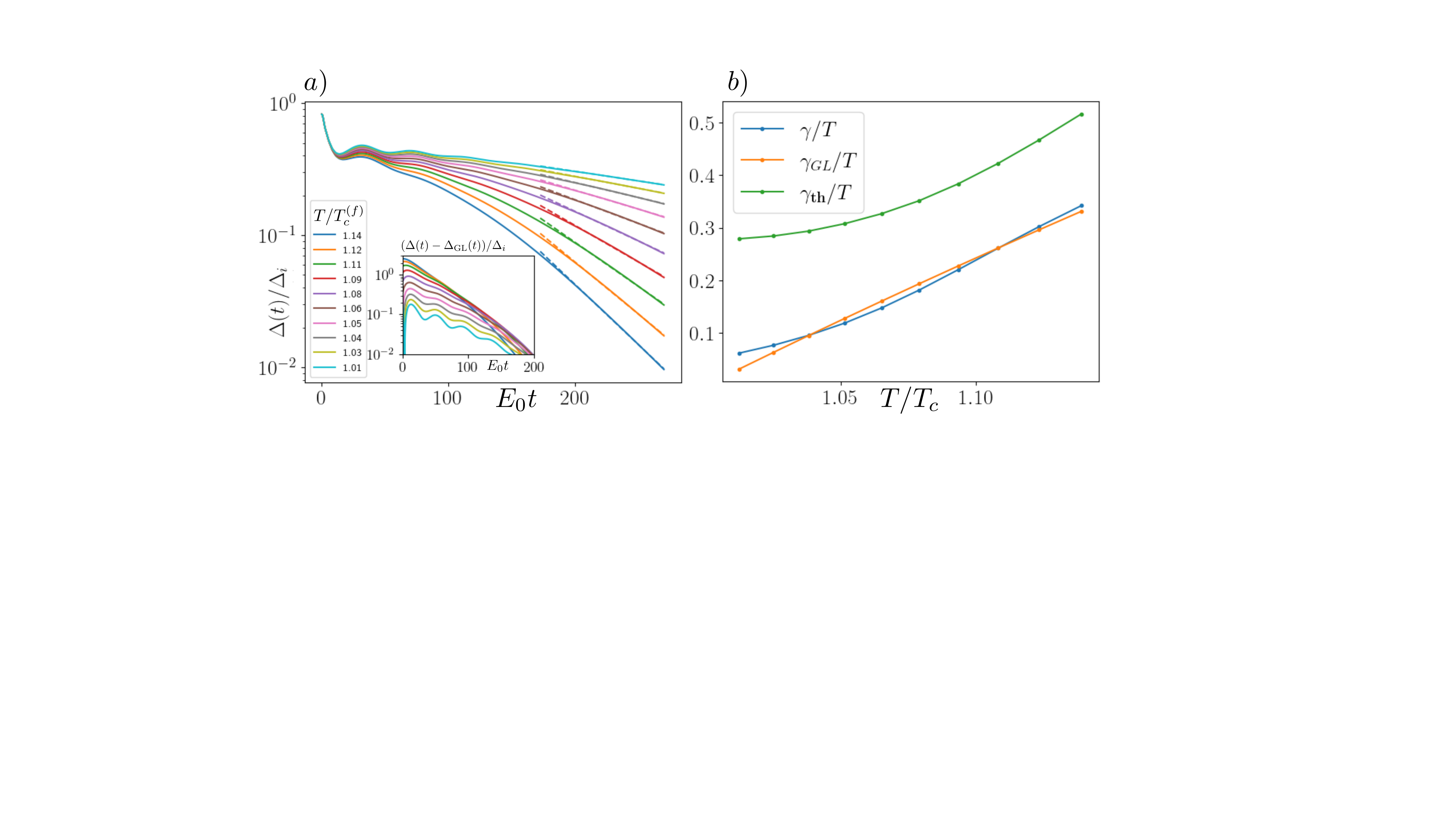}
\par\end{centering}
\caption{Quench with finite electron-phonon coupling assuming $\lambda_{i}=0.4,\lambda_{f}=0.33$,
$\lambda_{\text{el-ph}}=0.25$, $\bar{\omega}=0.125E_{0}$ at different
temperatures. a) Time-dependence of the order parameter at different phonon bath temperatures.  Exponential
fit at large times is shown in dashed lines. Inset shows the different
of the gap $\Delta\left(t\right)$ from the fitted long-time exponential
behavior $\Delta_{\text{GL}}\left(t\right)$. b) Double-exponential
fit parameters of the data. \useshortskip}

\label{Fig2}
\end{figure}

We now consider the quench of the BCS coupling constant in the presence of electron relaxation processes. Phonon coupling is kept constant 
during the quench. Having in mind possible relevance to experiment, we are particularly  interested in the regime where the final state of the system is normal. The numerical solution of Eqs.~(\ref{eq:G}, \ref{eq:sigma}) is shown in Fig.~\ref{Fig2}. In particular, we study the relaxation
of the order parameter at large times for different initial temperatures and assuming the electron-phonon coupling strength is $\lambda_{\text{el-ph}}\approx0.25$.
As expected, without the phonon coupling we recover the same results
as in Fig.~\ref{Fig1}. At stronger coupling strength we find the
the relaxation towards the equilibrium value ($\Delta_{\infty}=0$)
which follows an asymptotic behavior $\propto e^{-\gamma t}$. 
The relaxation rates $\gamma(T)$ are shown in Fig.~\ref{Fig2}~(b).
We find those to be in excellent agreement with the
the conventional time-dependent Ginzburg-Landau (GL) \citep{AHL95}, with $\gamma_{\text{GL}}=\frac{8}{\pi}T\log\frac{T}{T_{c}}$. We note that we do not expand the logarithm here in the vicinity of $T_{c}$ as we are interested in a wider range of temperatures. We also observe a minor deviation from the GL scaling at temperatures close to $T_{c}$, which may be due to phonons increasing quasiparticle relaxation processes further. 

The GL scaling implies that the distribution function is nearly equilibrium at large times. We numerically evaluate how
 $\Delta\left(t\right)$ approaches the GL scaling and find
another exponential behavior $\Delta\left(t\right)-\Delta_{\text{GL}}\left(t\right)$. As shown in Fig.~\ref{Fig2}~(b) the scaling is roughly exponential $\propto e^{-\gamma_{\text{th}}t}$
and we interpret the $\gamma_{\text{th}}$ as the thermalization rate
of the system. We also $\gamma_{\text{th}}$ is decreasing with lower
temperatures which we attribute to the overall gap being larger thereby
limiting the part of the phonon spectrum which participates in relaxation.

Let us now provide an intuitive picture behind the relaxation process.
When there is no phonon coupling, the state at each moment in time is a coherently excited gas of uncondensed Cooper pairs on top of the BdG vacuum (a small fraction of thermally excited quasiparticles do not play a dominant role in this regime). The oscillations of the order parameter as seen in Fig.~\ref{Fig2}~(b) can be interpreted
as a coherent bosonic oscillation showing periodic revivals (``re-condensation'' of this gas). For further insight, consider the density
matrix of the superconductor  $\hat{\rho}_{k}\left(t\right)=\langle\Psi_{k}\left(t\right)\otimes\Psi_{k}^{\dagger}\left(t\right)\rangle$.
Perform a unitary transformation into a comoving
frame, which diagonalizes the BdG Hamiltonian $H_{\text{BdG}}\left(t\right)$.
In this comoving frame, the effective Hamiltonian of the system has two terms:  $U^{\dagger}(t)H_{\text{BdG}}U(t)=\hat{h}_{\text{ad}}=\sum_{k}\sqrt{\Delta^{2}\left(t\right)+\xi_{k}^{2}}\Psi_{k}^{\dagger}\tau_{3}\Psi_{k}$ dynamics
and the  so-called adiabatic gauge potential
$\hat{h}_{G}=-i\partial_{t}\Delta\left(t\right)U^{\dagger}\tau_{1}\partial_{\Delta}U$
\citep{KSM17}. The first term represents a two-band system of the instantaneous BdG superconductor and the second term represents
an effective coherent driving which induces creation or
annihilation of uncondensed correlated Cooper pairs.
We note that in this basis the BCS gap equation is given by \useshortskip
\[
\Delta\left(t\right)=\lambda\int_{-E_{0}}^{E_{0}}d\xi_{k}\frac{\Delta\left(t\right)(1-2\tilde{\rho}_{k}^{(1,1)})+2\xi_{k}\text{Re}\tilde{\rho}_{k}^{(0,1)}}{2\sqrt{\Delta^{2}\left(t\right)+\xi_{k}^{2}}},
\]
\useshortskip where $\tilde{\rho}_{k}\left(t\right)=U^{\dagger}\left(t\right)\hat{\rho}_{k}\left(t\right)U\left(t\right)$.
This gap equation is different from some other non-equilibrium
scenarios as in case of e.g. the Eliashberg effect \citep{E72} which hinges on exciting fermionic quasiparticles. 

\begin{figure}
\includegraphics[scale=0.26]{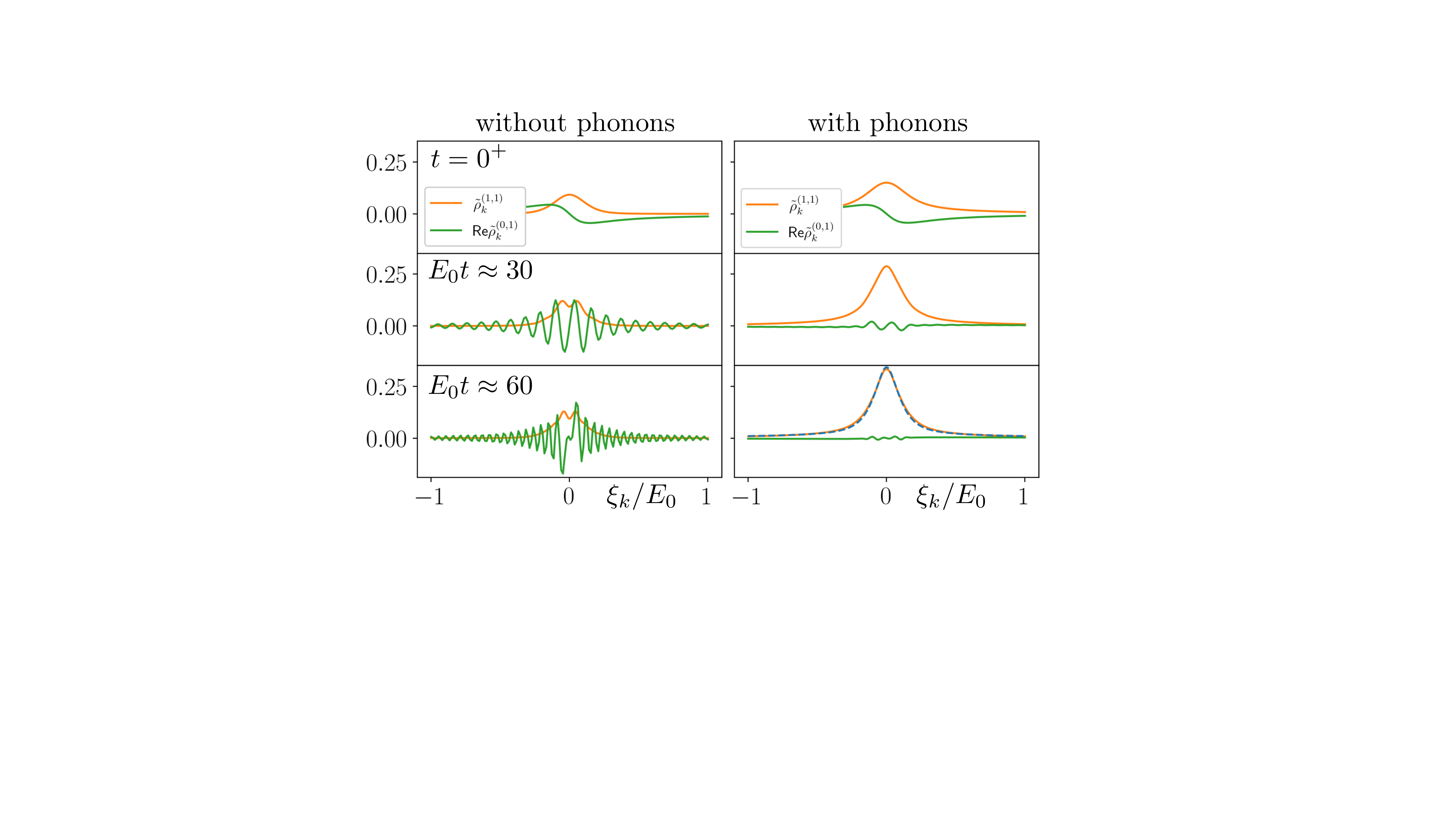}

\caption{Distribution of quasiparticles at different momenta $\xi_{k}$ in
non-equilibrium superconductor at different times $E_{0}t=0,30,60$
(from top to bottom) at $T/T_{c}^{(f)}\approx1.14$. These times roughly
correspond to $1/\gamma_{\text{th}}$ and $2/\gamma_{\text{th}}$
at this temperature. Left panel no phonon coupling, right panel: $\lambda_{\text{el-ph}}=0.25,$
$\bar{\omega}=0.125E_{0}$. Quench parameters are the same as in Fig.~\ref{Fig2}.
The blue dashed line corresponds to a thermal distribution at
temperature $0.85T$.}

\label{Fig3}
\end{figure}

We note here an important technical detail  that the different components of $\tilde{\rho}$ are not independent: the diagonal
terms contain contributions from both Cooper pairs and incoherent
electron excitations at any non-zero $T$. The latter be found by diagonalizing the matrix $\hat{\rho}_{k}$.
The result of numerical simulation without phonons is shown in Fig.~\ref{Fig3}~(left panel). We observe that uncondensed Cooper pairs persist at all times as expected in the absence of pair-breaking processes. The slow decay of the oscillatory behavior of the order parameter can be interpreted as dephasing of uncondensed Cooper pairs due to their oscillation
frequency mismatch. In the presence of phonons,  the emission and absorption of  phonons induce pair breaking \citep{CS77,KCL76}.
Phonons also renormalize the gap and its effect can be numerically
estimated at sufficiently large times from the off-diagonal term of
the retarded self-energy $\hat{\Sigma}_{\omega=0,T}^{R}\equiv\int d\tau\hat{\Sigma}_{\tau,T}^{R}$,
where the center-of-mass and relative times are defined as $T=(t+t')/2$
and $\tau=t-t'$. This contribution is usually small for our choice
of parameters. The distribution of quasiparticles affected by phonon
scattering is shown in Fig.~\ref{Fig3} (right panel). We observe
that at sufficiently long times, the distribution of electrons becomes
quasi-thermal and is nearly diagonal, corresponding to a nearly equilibrium electron configuration, which eventually approaches that at the phonon temperature. This justifies our interpretation of the $\gamma_{\text{th}}$
timescale as thermalization time in Fig.~\ref{Fig2}. In conclusion,
we note that our findings imply that the thermalization stage in our
setup cannot be adequately described by the conventional kinetic equation
approach to superconductors, where the distribution function is assumed
to be diagonal \citep{KCL76,CS77,E72}. Proper analytical understanding
of dynamics of uncondensed pairs necessarily requires the consideration
of a more general matrix-valued distribution \citep{K23} function.

\paragraph{Conclusions}

In this work, we considered the dynamics and thermalization of a
superconductor where the pairing interaction is rapidly quenched from $\lambda_i$ to a lower value $\lambda_f < \lambda_i$. This may qualitatively corresponds to the regime in light-induced superconductors after the irradiation is turned off.  First, we focused on a model without pair-breaking processes
 where the dynamics of the order parameter is dominated by coherent bosonic oscillations, which correspond to the integrable regime considered previously for initial zero-temperature states. We explored this regime for a wider range of initial conditions, including thermal initial states. Surprisingly, we found that the order
parameter can relax to a finite steady-state value even if the initial temperature exceeds the critical temperature corresponding to the final value of the interaction strength: $T_c(\lambda_i) > T_c(\lambda_f)$. We then explored the relaxation in the
presence of a phonon bath, which leads to a crossover from integrable to ergodic dynamics and  thermalization, dominated by fermionic mechamism due to pair-breaking processes. Specifically, we demonstrate that at sufficiently long times, the electron distribution becomes nearly diagonal and quasi-thermal. Our approach lays the ground for numerous future research opportunities including more microscopic studies of light-induced superconductivity, Floquet-enhanced pairing under periodic driving, and spatially inhomogeneous non-equilibrium superconductivity.

\paragraph{Acknowledgments}
This work was supported by the U.S. Department of Energy, Office of Science, Basic Energy Sciences under Award No. DE-SC0001911. V.G. acknowledges useful discussions with Andrea Cavalleri, Andy Millis, and Vadim Oganesyan and thanks the Flatiron Institute and the Simons Foundations for hospitality and support during the early stages of this project.
\bibliographystyle{apsrev4-2}
\bibliography{biblio}

\end{document}